\documentclass[12pt]{article}
\usepackage{graphicx}
\usepackage{hyperref}
\usepackage{caption}

\newcommand\pubnumber{Article 27 in eConf C1304143}
\newcommand\pubdate{\today}

\newcommand\Bt{\rule[-1.0ex]{0pt}{0pt}}

\newcommand\Tt{\rule{0pt}{2.5ex}}

\def\napoli{$^{1}$E\"otv\"os University, Budapest, Hungary,
$^{2}$MTA CSFK Konkoly Observatory, Budapest, Hungary,
$^{3}$National University of Public Service, Budapest, Hungary}

\def\Title#1{\begin{center} {\Large #1 } \end{center}}
\def\Author#1{\begin{center}{ \sc #1} \end{center}}
\def\Address#1{\begin{center}{ \it #1} \end{center}}

\newcommand\pubblock{\rightline{\begin{tabular}{l} \pubnumber\\
         \pubdate  \end{tabular}}}
\newenvironment{Abstract}{\begin{quotation}  }{\end{quotation}}
\newenvironment{Presented}{\begin{quotation} \begin{center} 
             PRESENTED AT\end{center}\bigskip 
      \begin{center}\begin{large}}{\end{large}\end{center} \end{quotation}}
\def\Acknowledgements{\bigskip  \bigskip \begin{center} \begin{large}
             \bf ACKNOWLEDGEMENTS \end{large}\end{center}}

\def\beq{\begin{equation}}
\def\eeq#1{\label{#1}\end{equation}}
\def\eeqn{\end{equation}}

\def\beqa{\begin{eqnarray}}
\def\eeqa#1{\label{#1}\end{eqnarray}}
\def\eeqan{\end{eqnarray}}

\let\bar=\overbar

\def\Dslash{\not{\hbox{\kern-4pt $D$}}}
\def\dslash{\not{\hbox{\kern-2pt $\del$}}}

\def\msb{{\bar{\ssstyle M \kern -1pt S}}}

\begin{document}
\begin{titlepage}
\pubblock

\vfill
\Title{Statistical analysis of the prompt and afterglow emission of the three groups of gamma-ray bursts}
\vfill
\Author{J. K\'obori$^{1}$, Z. Bagoly$^{1}$, L. G. Bal\'azs$^{1,2}$ and I. Horv\'ath$^{3}$}
\Address{\napoli}
\vfill
\begin{Abstract}
We investigated the main prompt and afterglow emission parameters of gamma-ray
bursts detected by the Burst Alert Telescope (BAT) and X-Ray Telescope installed on the Swift satellite. 
Our aim was to look for differences or connections 
between the different types of gamma-ray bursts, so we compared the BAT fluences, 1-sec peak
photon fluxes, photon indices, XRT early fluxes, initial temporal decay and spectral
indices. We found that there might be a connection between the XRT initial decay 
index and XRT early flux/BAT photon index. Using statistical tools we also determined that beside
the duration and hardness ratios, the means of the $\gamma$- and X-ray--fluences and the $\gamma$-ray photon index differ significantly
between the three types of bursts.
\end{Abstract}
\vfill
\begin{Presented}
Huntsville Gamma Ray Burst Symposium, GRB 2013,
Nashville, Tenesse, USA
\end{Presented}
\vfill
\end{titlepage}
\def\thefootnote{\fnsymbol{footnote}}
\setcounter{footnote}{0}

\section{Introduction}

Gamma-ray bursts (GRBs), based on their duration and hardness ratios, 
can be divided into three groups: short (SB), intermediate (IB) and long duration bursts (LB).
While the progenitors of SBs and LBs are very likely different, the IBs and LBs show similar
features in their prompt and afterglow emission. So far, many authors (e.g. de Ugarte Postigo et al. 2010, \cite{postigo}) 
investigated whether the observed differences or similarities in various parameters concerning the IBs and LBs are significant or not.
Also, the presence of the intermediate group can be a result of observational and/or instrumental effects. An important question is that
apart from the duration and hardness ratio parameters are there any other quantities which possess significant differences in their measured values.

\section{Data reduction - methods}
The data was downloaded from the Heasarc Archive
System (\cite{grbtab}), the sample consists of GRBs up to 27 March
2013, but we only included those bursts which have
measured BAT fluences, 1-sec peak photon fluxes,
photon indices, XRT early fluxes, initial temporal decay and spectral indices. The sample
consists of 317 bursts: 28 short, 55 intermediate
 and 234 long duration bursts. The classification was based on the $T_{90}$ 
parameter, where mean
values for the different groups are: 0.47 s, 13.7 s and
50.0 s for the SBs, IBs and LBs, respectively (\cite{veres}).
In order to determine the significances in the differences between the SBs, IBs and LBs regarding the above parameters, linear
discriminant analysis was carried out on the sample. The details of this method can be read in Section 6.

\section{BAT properties}
\begin{figure}
\begin{minipage}[b]{0.3\linewidth}
\centering
\includegraphics[scale = 0.45]{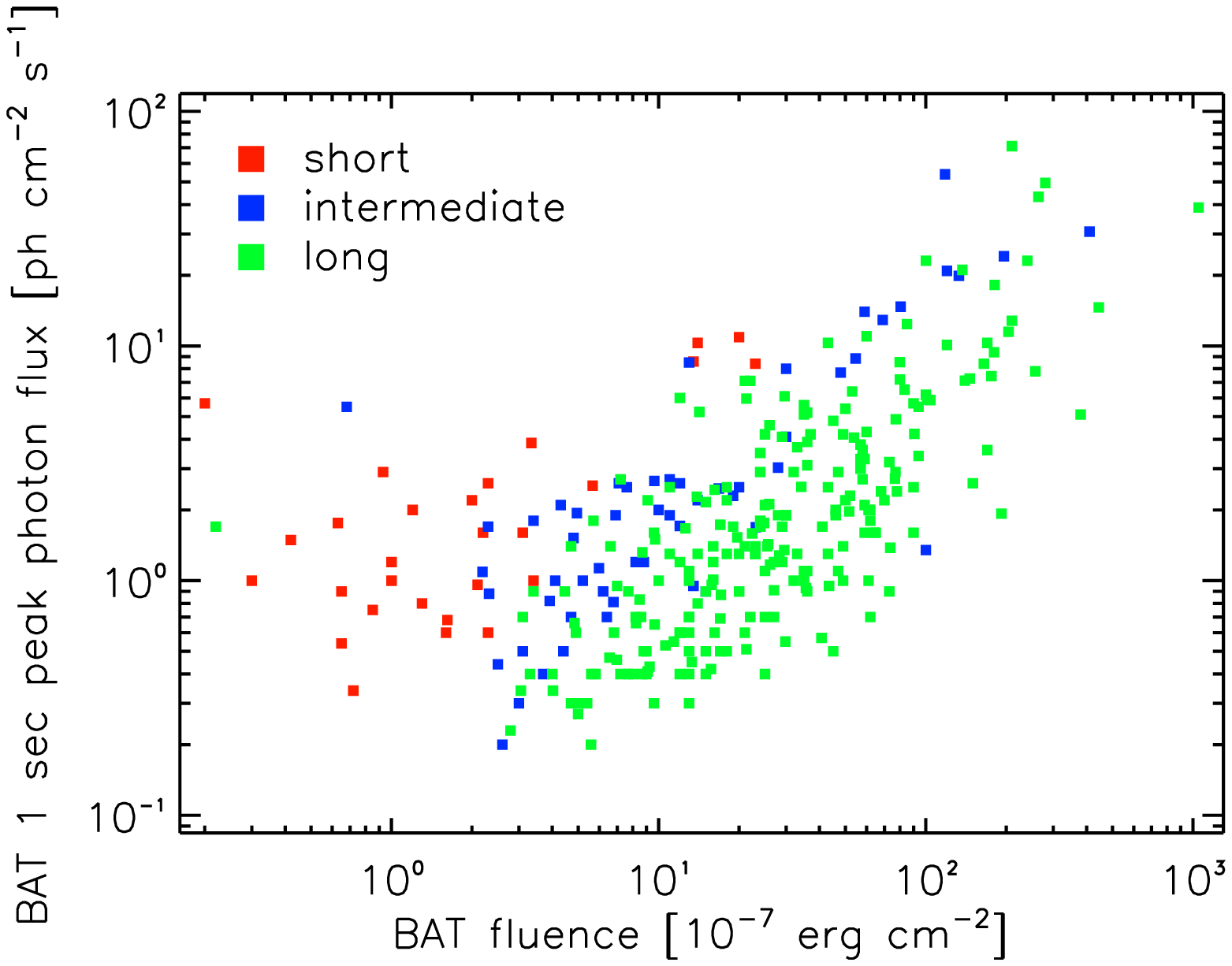}
\end{minipage}
\hspace{3cm}
\begin{minipage}[b]{0.3\linewidth}
\centering
\includegraphics[scale = 0.45]{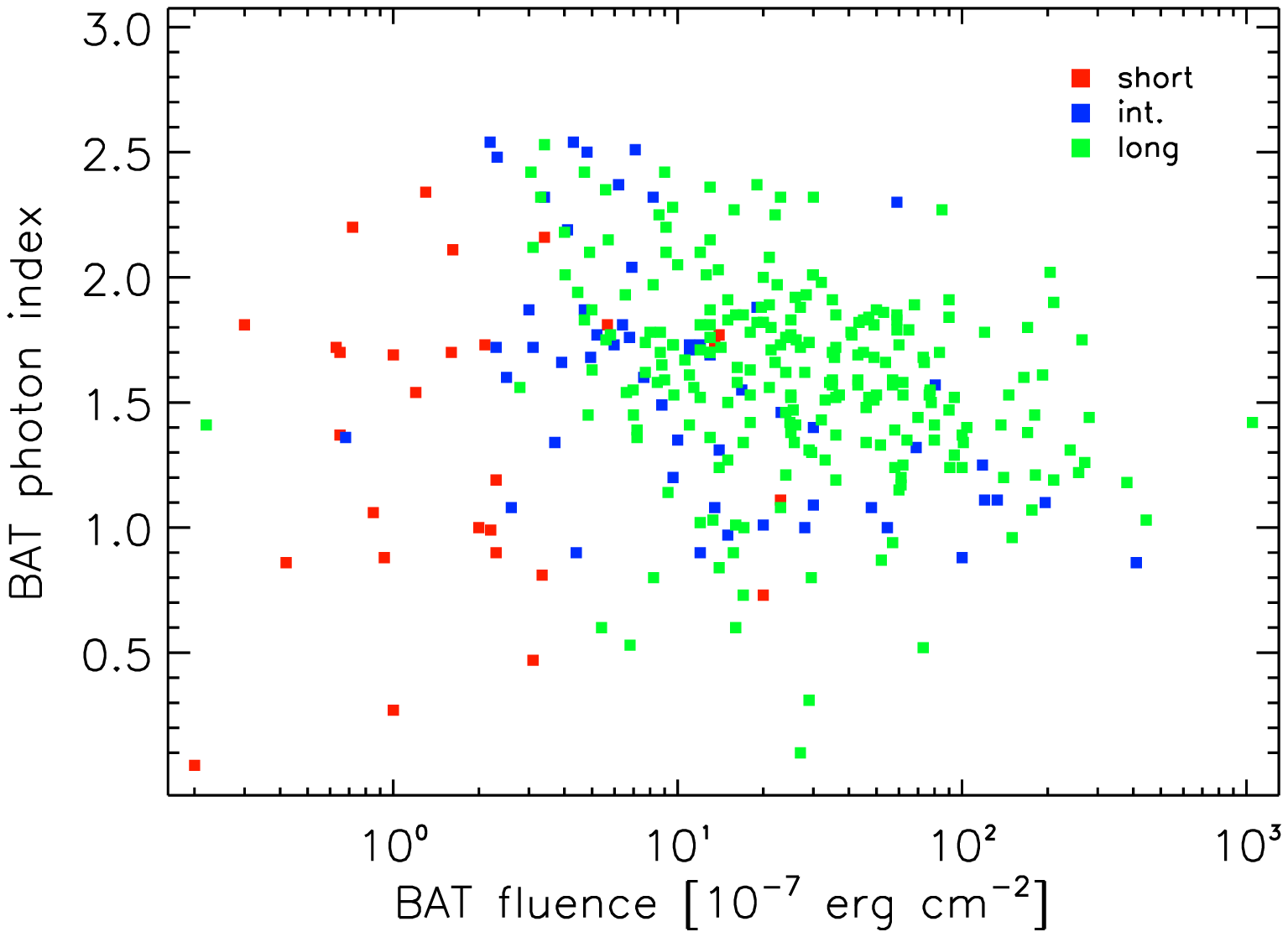}
\end{minipage}
\caption{On the left side the fluence and the 1-s peak photon flux are plotted. It can be observed that
the three types of bursts form three almost distinctive area, although, the long and intermediate bursts share a relatively wide interval.
The right panel shows that the BAT photon index does not seem to depend on the fluence. Nevertheless, the former quantity has an upper bound,
which is practically the same for all types of bursts.}
\label{fig1}
\end{figure}

First, we compared the main BAT measured parameters: the fluence, the 1-sec peak photon flux and the
photon index. As we can see on the Fig. \ref{fig1} (left side) the
three types of bursts separate quite well on the fluence and peak photon flux plane.
Regarding the photon index and the peak photon
flux variables, the three groups almost entirily
overlap, however, the short and long bursts, according to the distinct distribution on the fluence -- peak 
photon flux plane, do not show any similarity between
the fluence and photon index parameters (see Fig. \ref{fig1}, right side).

\section{XRT properties}
\begin{figure}[h!]
\begin{minipage}[b]{0.3\linewidth}
\centering
\includegraphics[scale = 0.45]{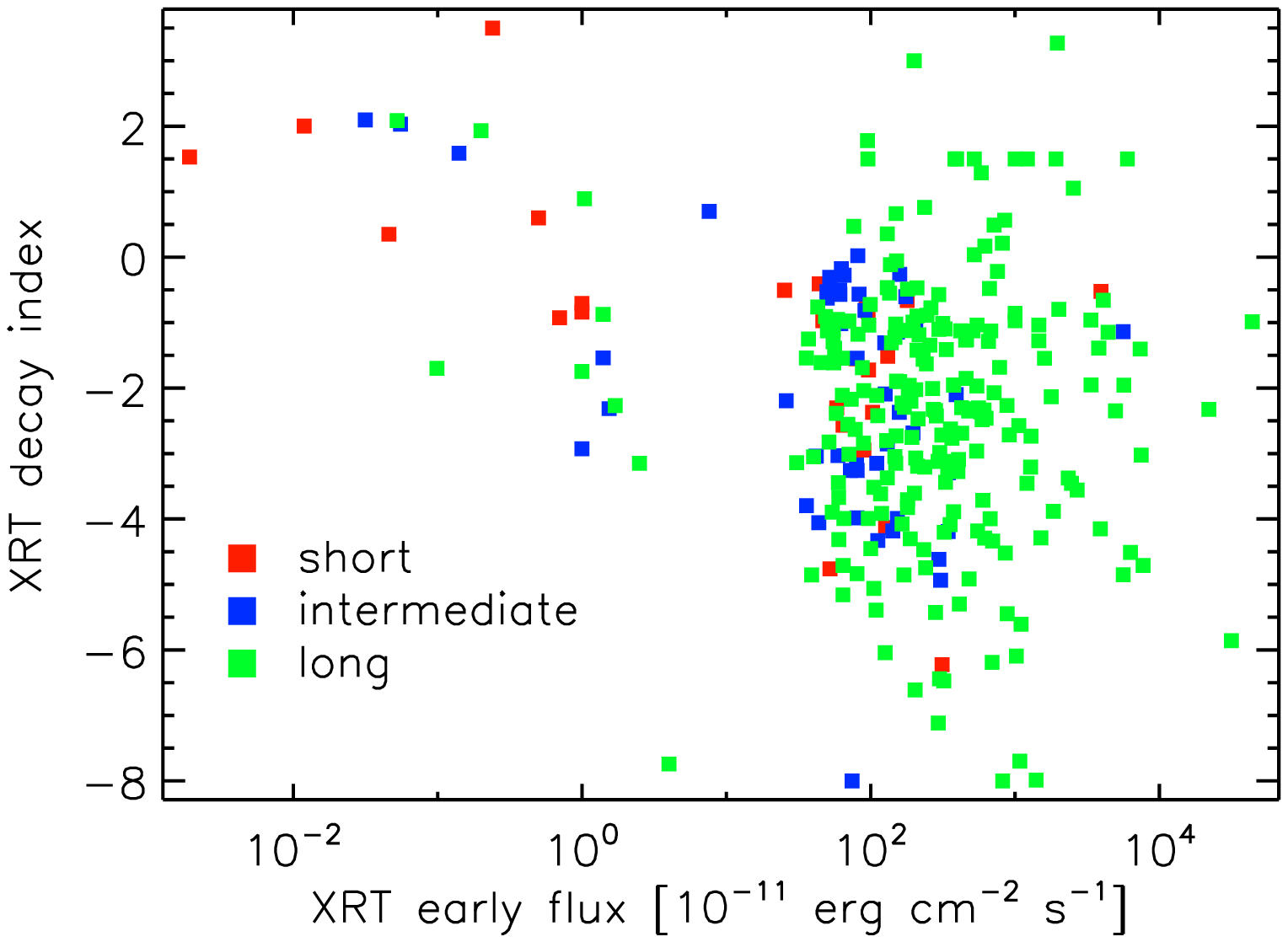}
\end{minipage}
\hspace{3cm}
\begin{minipage}[b]{0.3\linewidth}
\centering
\includegraphics[scale = 0.45]{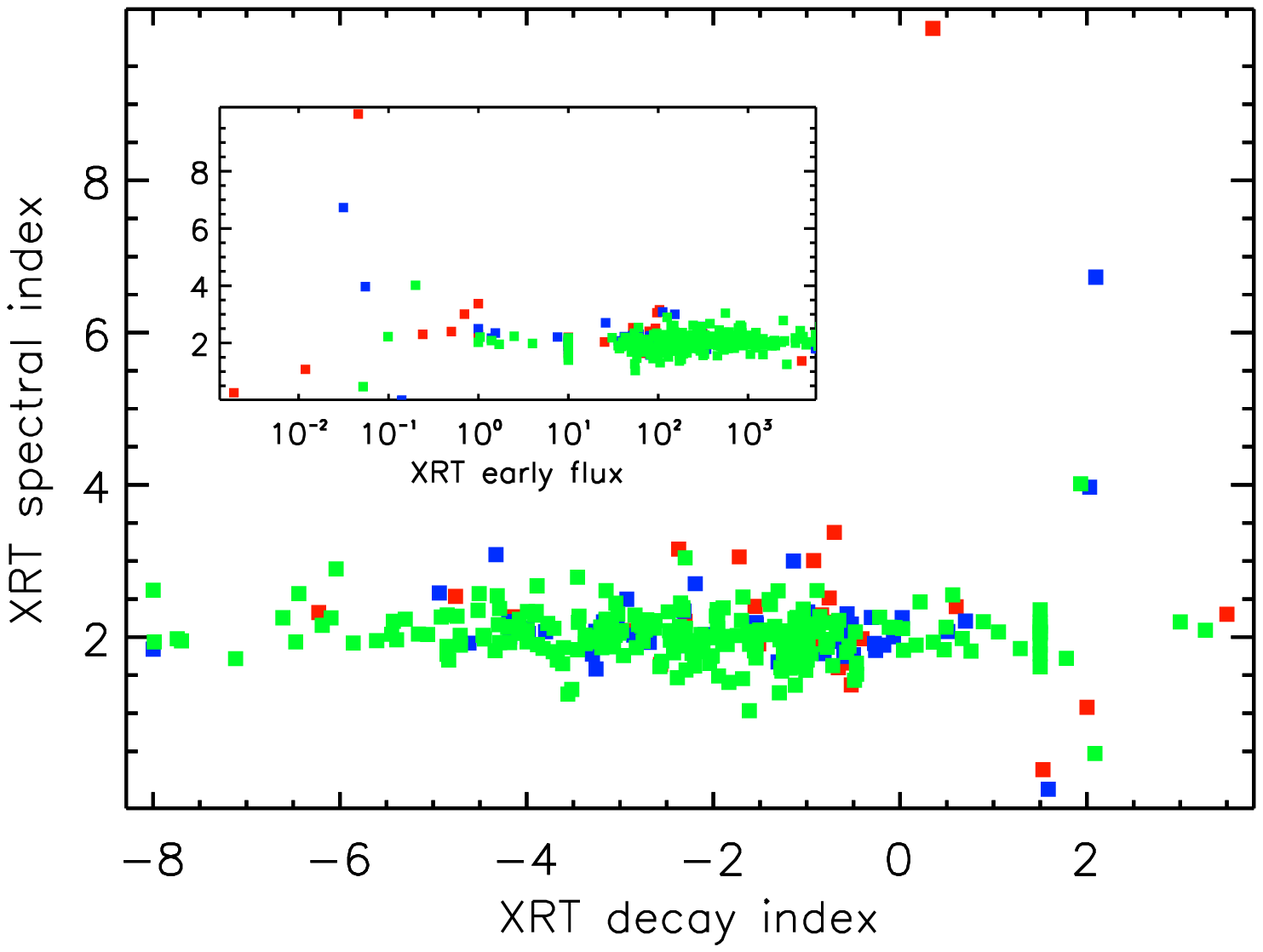}
\end{minipage}
\caption{Left panel: An interesting result found regarding the XRT early flux and decay index is that 
the decay slope converges to 2 as the early flux gets lower for all bursts. In addition, two
distinct group of early flux can be observed. However, we should note that the quality and reliability of 
numbers of this quantity in the Swift GRB Table is quite low (P. Evans, private communication).
The right panel shows the XRT spectral index dependence on the decay index and early flux. No correlation
found between them.}
\label{fig2}
\end{figure}

We also compared the XRT afterglow features. An
interesting result we found that X-ray decay index
seems to depend on the XRT early flux (see Fig. \ref{fig2}, left panel). More precisely, in the cases of
the LBs the higher the early flux, the broader
the decay index range, however, IBs
show quite similar behaviour, but we should notice
that, on one hand, the latter type of bursts tend to
have lower early fluxes, and on the other hand the
trend can be a result of the low number of short and
IBs. Contrary to the previous correlation the XRT spectral index does not depend on the decay index and early flux 
(Fig. \ref{fig2}, right panel).

\section{BAT vs. XRT features}
\begin{figure}[h!]
\begin{minipage}[b]{0.3\linewidth}
\centering
\includegraphics[scale = 0.45]{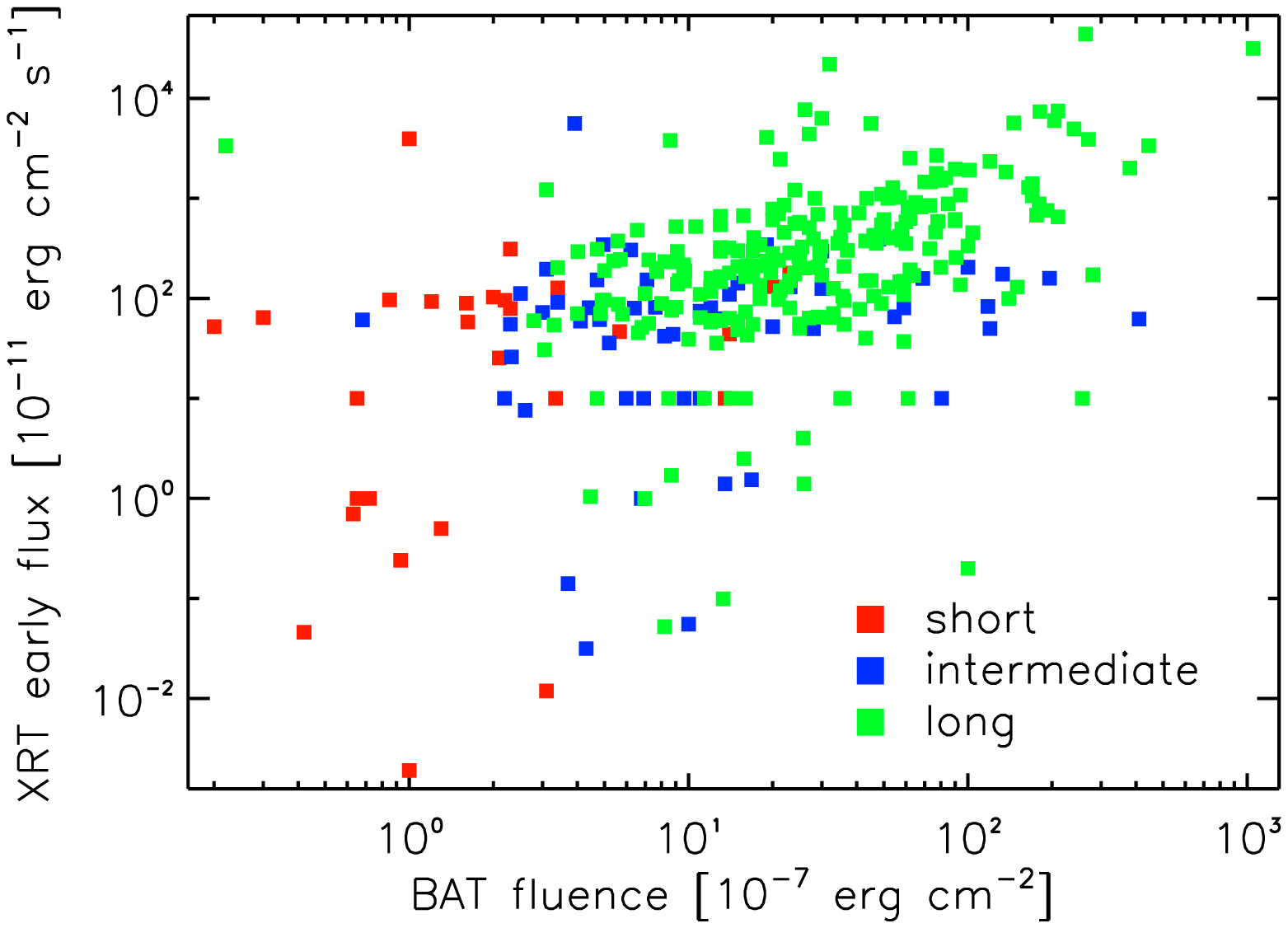}
\end{minipage}
\hspace{3cm}
\begin{minipage}[b]{0.3\linewidth}
\centering
\includegraphics[scale = 0.45]{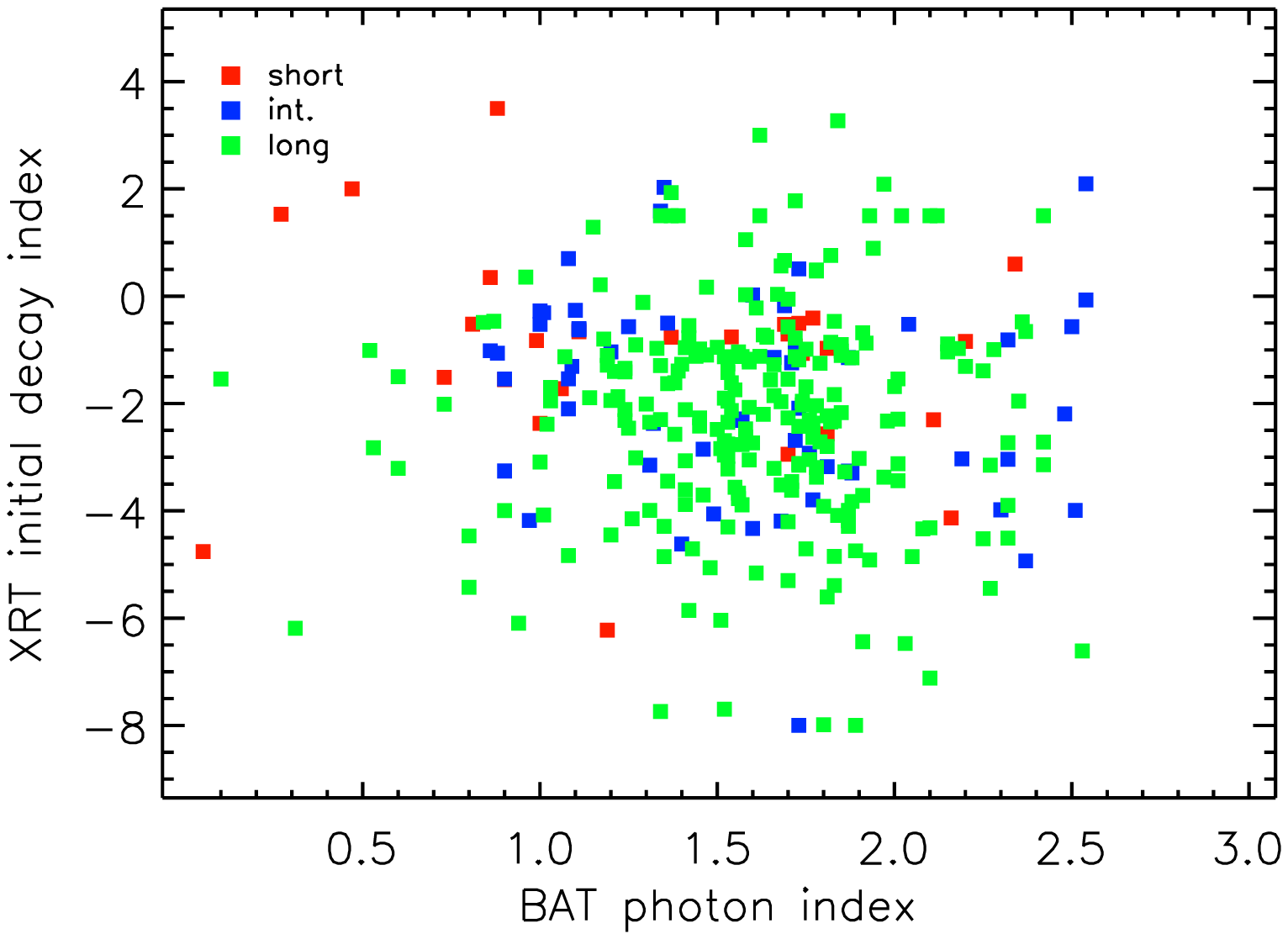}
\end{minipage}
\caption{On the left side it is seen that the XRT early flux for SBs and IBs have an upper limit, while for the LBs
it can take any value above the mentioned upper bound. On the right side we can see the XRT initial decay index plotted against the
BAT photon index. It is interesting that the decay index is not independent from the $\gamma$-photon index, more precisely,
softer $\gamma$-ray spectra correspond to narrower decay index range.}
\label{fig3}
\end{figure}

A more interesting issue whether the BAT and XRT
properties correlate or not. In order to test this we correlated all of the BAT and XRT features against each other and
among them we found two connections.
The XRT early flux for the IBs seems to have an 
upper limit around $(10^2 - 10^3) \times 10^{-11}$ erg/cm$^2$/s (except one outlier), while for the LBs this quantity shows a continous 
distribution (Fig. \ref{fig3}, left side). The other remarkable result is dependence of the XRT initial decay index on the BAT photon index (Fig. 
\ref{fig3}).
In the cases of the LBs there is a weak correlation
between them, as the BAT photon index gets lower,
the XRT decay index tends to converge to the value of
$\sim$(-2). The IBs and SBs do not show such properties.

\section{Statistical method}
The Discriminant Analysis (DA) is a statistical technique 
which allows us, based on a priori classification, to find the linear combination of the explanatory
variables which characterize or separates two or more
classes.
Using the SPSS statistical program package (\cite{spss}) we carried out such an analisys on the sample.
The purpose of DA is to estimate the
relationship between a single categorical dependent variable and
a set of quantitative independent variables.
DA involves deriving a variate, the linear
combination of the two (or more) independent variables that will
discriminate best between defined groups. The linear combination
for a discriminant analysis, also known as the discriminant
function is derived from an equation that takes the following
form (there are altogether $k − 1$ discriminant functions, where $k$ is the number of classes):
\begin{equation}
y = n_1x_1 + n_2x_2 + ... + n_px_p, \hspace{1cm}\textrm{where} \hspace{1cm}\sum_{j=1}^p n_j^2= 1
\end{equation}
In our case we have 3 groups, so we are looking for 2 discriminant functions (variables). The
results of this analysis are shown in Table \ref{t1} and \ref{t2}.
\begin{table}[ht]
\begin{center}
\begin{minipage}[b]{0.3\linewidth}\centering
\caption{Tests of equality of group means}
\begin{tabular*}{1.3\textwidth}{@{\extracolsep{\fill}}lcr}
\noalign{\hrule height 1.5pt}
 \Bt \Tt& F  \Bt \Tt& Sig. \\
\noalign{\hrule height 1.5pt}
\bf{logFlu}  & 56.387& \Bt \Tt \bf{0.000} \\
logPeak &   0.089& \Bt \Tt 0.915 \\
\bf{logXflu} & 18.525&   \Bt \Tt \bf{0.000} \\
\bf{Pind}    &  8.002&  \Bt \Tt \bf{0.000} \\
Xdec    &  1.186&   \Bt \Tt0.307 \\
\bf{Xsp}   &  5.881&  \Bt \Tt\bf{0.003} \\
\noalign{\hrule height 1.5pt}
\label{t1}
\end{tabular*}
\end{minipage}
\hspace{1.5cm}
\begin{minipage}[b]{0.33\linewidth}
\caption{Structure matrix}
\vspace{1mm}
\centering
\begin{tabular*}{1.\textwidth}{@{\extracolsep{\fill}}lcr}
\noalign{\hrule height 1.5pt}
& Func. 1 & Func. \Bt \Tt 1 \\
\noalign{\hrule height 1.5pt}
logFlu   & 0.694 \Bt \Tt   &0.217 \\
logPeak  &0.004\Bt \Tt	&0.092 \\
logXflu  &0.376\Bt \Tt	&0.454 \\
Pind     &0.147\Bt \Tt	&-0.735 \\
Xdec   &-0.096\Bt \Tt	&0.107 \\
Xsp    &-0.224\Bt \Tt  &  0.066 \\
\noalign{\hrule height 1.5pt}
\label{t2}
\end{tabular*}
\end{minipage}
\caption*{In Table 1. one can see the significances of the differences in the group means. $F$ is
the test variable which characterizes the ratio of the variances between and within the groups.
Bold face indicates the cases where the difference is significant. In Table 2. the members of the so-called Structure matrix can be read. These coefficient denote that in the discriminant functions 1 and 2 
which variables dominate. In our case Function 1 is dominated by $\gamma-$ and X-ray - fluences, while Function 2 is governed by the photon index.}
\end{center}
\end{table}

\section{Discussion and conclusion}
In this article we examined whether the main BAT and XRT properties correlate or not. 
Apparently, the XRT early flux and early decay index are connected, as the former one decreases, the latter one
converges to the value of 2. Since the positive value means a rising afterglow (i.e. $F \propto t^{\alpha}$), one reason for this behaviour could be
that the Swift simply starts to observe these bursts earlier, so the time difference between the peak and first observation is greater
for bursts with lower XRT early flux. The XRT early flux also seems to depend on the BAT photon index, although, in this case
it converges to $\sim$ (-2), while the photon index decreases. This issue is a more problematic one, since the $\gamma$- and X-ray emissions
could be physically connected and/or detached, so our plan is to investigate further this problem. 
We also determined which 
prompt and afterglow parameters can be used to distinguish the three types of bursts. Using the Linear Discriminant Analysis
statistical tool, we come to the conclusion, that the 
$\gamma-$ and X-ray fluences as well as the BAT photon index are the most important quantities, beside the
duration and hardness ratios, which separate mostly the SBs, IBs and LBs.

\Acknowledgements
This work was supported by the Hungarian OTKA-77795 grant.
This research has made use of data obtained through
the High Energy Astrophysics Science Archive Research Center Online Service, provided by the
NASA/Goddard Space Flight Center.
I acknowledge Doctorate School of Physics, E\"otv\"os University, for supporting my participation on
the Huntsville Gamma Ray Burst Symposium, GRB 2013, Nashville, Tennessee, USA.

\end{document}